\documentclass[sigconf]{acmart}
\def\BibTeX{{\rm B\kern-.05em{\sc i\kern-.025em b}\kern-.08emT\kern-.1667em\lower.7ex\hbox{E}\kern-.125emX}}

\AtBeginDocument{%
  \providecommand\BibTeX{{%
    \normalfont B\kern-0.5em{\scshape i\kern-0.25em b}\kern-0.8em\TeX}}}

\copyrightyear{}
\acmYear{}
\setcopyright{none}
\acmConference[CIKM Workshop 2019]{Workshop}
\begin{document}

%
% The "title" command has an optional parameter, allowing the author to define a "short title" to be used in page headers.
\title{UBER-GNN: A User-Based Embeddings Recommendation based on Graph Neural Networks }

%
% The "author" command and its associated commands are used to define the authors and their affiliations.
% Of note is the shared affiliation of the first two authors, and the "authornote" and "authornotemark" commands
% used to denote shared contribution to the research.
%\author{ $\quad$   }
%\authornote{$\quad$ }
%\email{$\quad$ }
%\orcid{$\quad$ }
%\author{}
%%\authornotemark[1]
%\email{}
%\affiliation{%
%  \institution{}
%  \streetaddress{}
%  \city{}
%  \state{}
%  \postcode{}
%}
%
%
%\author{}
%\affiliation{%
%  \institution{}
%  \city{}
%  \country{}
%}
%
%\author{}
%\affiliation{%
% \institution{}
% \streetaddress{}
% \city{}
% \state{}
% \country{}}
\author{Bo Huang}
\affiliation{%
  \institution{Ping An Technology (Shenzhen) Co., \\Ltd}
  \city{Shanghai, China}
}
\email{huangbo098@pingan.com.cn}

\author{Ye Bi}
\affiliation{%
  \institution{Ping An Technology (Shenzhen) Co., \\Ltd}
  \city{Shanghai, China}
}
\email{biye645@pingan.com.cn}

\author{Zhenyu Wu}
\affiliation{%
  \institution{Ping An Technology (Shenzhen) Co., \\Ltd}
  \city{Shanghai, China}
}
\email{wuzhenyu447@pingan.com.cn}

\author{Jianming Wang}
\affiliation{%
  \institution{Ping An Technology (Shenzhen) Co., \\Ltd}
  \city{Shanghai, China}
}
\email{wangjianming888@pingan.com.cn}

\author{Jing Xiao}
\affiliation{%
  \institution{Ping An Technology (Shenzhen) Co., \\Ltd}
  \city{Shanghai, China}
}
\email{xiaojing661@pingan.com.cn}

%
% By default, the full list of authors will be used in the page headers. Often, this list is too long, and will overlap
% other information printed in the page headers. This command allows the author to define a more concise list
% of authors' names for this purpose.
\renewcommand{\shortauthors}{}

%
% The abstract is a short summary of the work to be presented in the article.
\begin{abstract}
The problem of session-based recommendation aims to predict user next actions based on session histories.
Previous methods models session histories into sequences and estimate user latent features by RNN and GNN methods to make recommendations.
However under massive-scale and complicated financial recommendation scenarios with both virtual and real commodities , such methods are not sufficient to represent accurate user latent features and neglect the long-term characteristics of users.

To take long-term preference and dynamic interests into account, we propose a novel method, i.e. User-Based Embeddings Recommendation with Graph Neural Network, UBER-GNN for brevity.
UBER-GNN takes advantage of structured data to generate long-term user preferences, and transfers session sequences into graphs to generate graph-based dynamic interests.
The final user latent feature is then represented as the composition of the long-term preferences and the dynamic interests using attention mechanism.

Extensive experiments conducted on real Ping An scenario show that UBER-GNN outperforms the state-of-the-art session-based recommendation methods.
\end{abstract}

%
% The code below is generated by the tool at http://dl.acm.org/ccs.cfm.
% Please copy and paste the code instead of the example below.
%
%\begin{CCSXML}
%<ccs2012>
% <concept>
%  <concept_id>10010520.10010553.10010562</concept_id>
%  <concept_desc>Computer systems organization~Embedded systems</concept_desc>
%  <concept_significance>500</concept_significance>
% </concept>
% <concept>
%  <concept_id>10010520.10010575.10010755</concept_id>
%  <concept_desc>Computer systems organization~Redundancy</concept_desc>
%  <concept_significance>300</concept_significance>
% </concept>
% <concept>
%  <concept_id>10010520.10010553.10010554</concept_id>
%  <concept_desc>Computer systems organization~Robotics</concept_desc>
%  <concept_significance>100</concept_significance>
% </concept>
% <concept>
%  <concept_id>10003033.10003083.10003095</concept_id>
%  <concept_desc>Networks~Network reliability</concept_desc>
%  <concept_significance>100</concept_significance>
% </concept>
%</ccs2012>
%\end{CCSXML}

%\ccsdesc[500]{Computer systems organization~Embedded systems}
%\ccsdesc[300]{Computer systems organization~Redundancy}
%\ccsdesc{Computer systems organization~Robotics}
%\ccsdesc[100]{Networks~Network reliability}

%
% Keywords. The author(s) should pick words that accurately describe the work being
% presented. Separate the keywords with commas.

%
% A "teaser" image appears between the author and affiliation information and the body
% of the document, and typically spans the page.

%
% This command processes the author and affiliation and title information and builds
% the first part of the formatted document.
\settopmatter{printacmref=false}

\maketitle
\section{Introduction}
\begin{figure}
  \centering
  \includegraphics[scale=0.15]{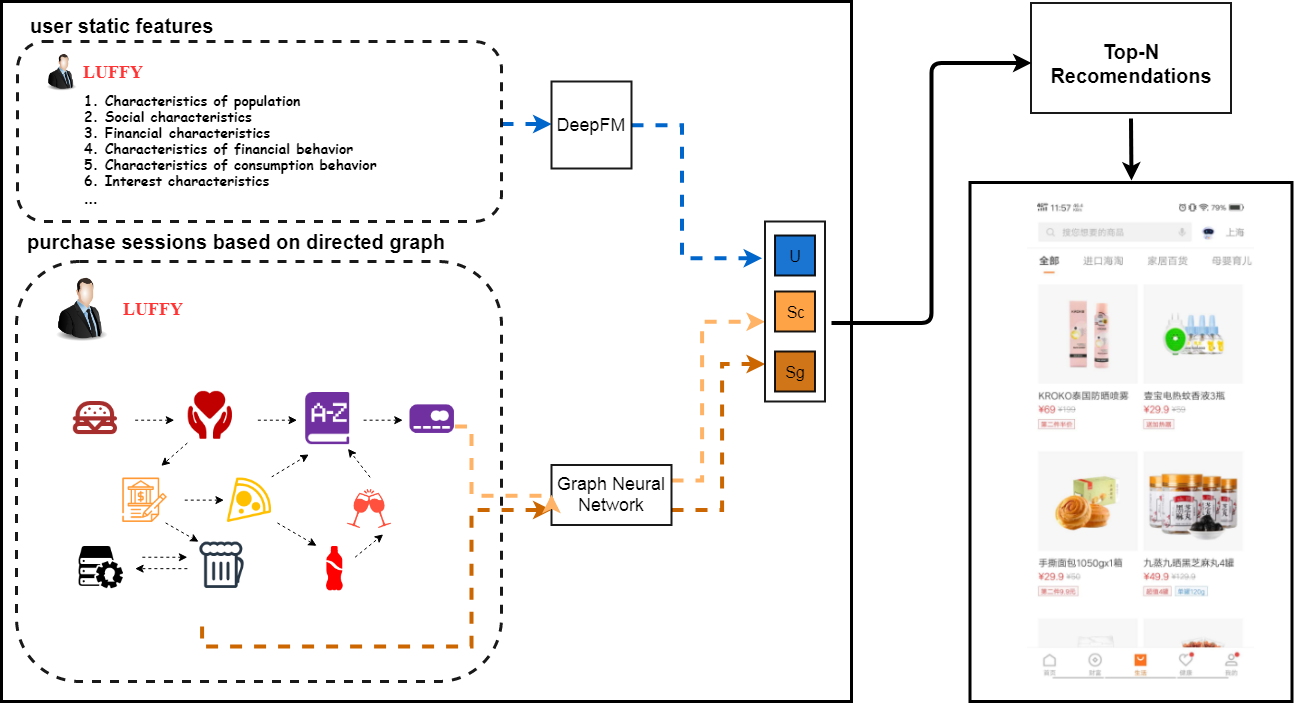}
  \caption{Recommendation scenario of Ping An Jinguanjia: it generates user-context based embeddings of user's characteristics and preferences via DeepFM. Meanwhile, it generates
  session based embbedings of user's purchase histories via attention network of GNN. Then though the online well-trained model, it updates the probabilities of Top-N recommendations and pushes selected commodities to APP-page.}
\end{figure}

In financial service category, Ping An \textbf{Jinguanjia}, JD Finance, Ant Fortune (Ant Financial) are top 3 applications on Mobile. As a comprehensive app ensembles diverse functions such as insurance purchases, investment services, Health consulting and so on, Jinguanjia also provides a typical E-Commerce platform for online shopping, serving over 15,000,000 users per month.

Besides Jinguanjia, with the high-speed development of China's Mobile Internet, more and more online transactions are driven by intelligent and efficient recommender algorithms. E-Commerce platforms like Alibaba,
searching and advertising platforms like Baidu, and O2O lifestyle
service platforms like Meituan, have their our recommender algorithm systems based on methods like association rules, Machine Learning(eg. collaborative filtering).
Moreover, with the rapid improvement of computers in recent years, algorithms like DNNs have made breakthroughs in aspects like Computer Visions and others, more and more domestic E-Commerce and advertising platforms have been contributing new explorations and innovations. For instance,
In \cite{WideandDeep}, members from Google Group explored a novel recommendation framework called Wide\&Deep for jointly training feed-forward neural networks with embedding and linear model with feature transformations; In\cite{DBLP:journals/corr/abs-1901-08907} , Meituan-Dianping Group advocated a Multi-task feature learning approach for knowledge graph enhanced recommendation, which presented a end-to-end deep KG network; In \cite{DBLP:journals/corr/abs-1711-06632}, the group from Alibaba Group proposed a novel recommendation framework through attention-based DNN in consist of the heterogeneous behaviors of users.
\begin{figure*}
  \centering
  \includegraphics[scale=0.3]{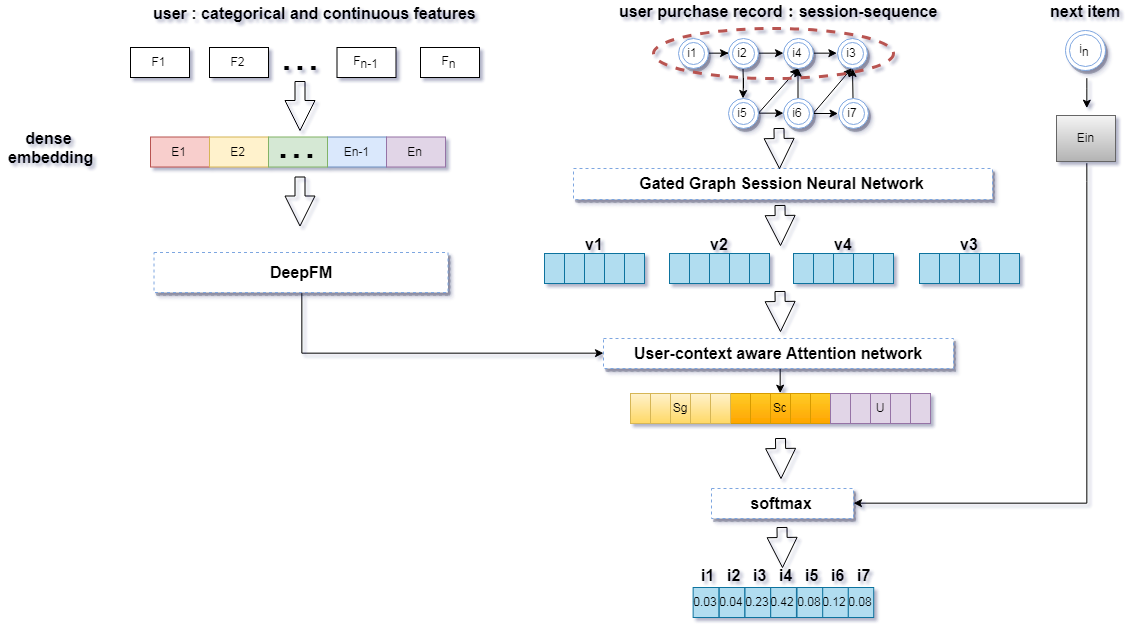}
  \caption{The architecture of UBER-GNN: On the left,we extract user-context based embeddings from the static features of users via DeepFM; On the right, we represent purchase session records as session graphs and generate
  session embeddings after fed into gated graph session neural network. Then we make the combination of  latent user embeddings and graph-based latent item embeddings via attention network. At last, we predict the probability for next-purchase one for each session.    }
\end{figure*}

Nevertheless, Deep Learning framework is not effective enough on the E-Commerce platform of \textbf{Jinguanjia}, where users can select various items more than hundred
thousands types, including virtual financial commodities like short-term medical insurance, physical examination services, and real commodities like fruits, snacks, etc.
Considering the mixture recommendation with virtual and real commodities, different from traditional E-Commerce only with real commodities, it is more complicated to predict the next-purchase item of users.  During the practice, the prediction methods of CTR (click-through rate) like DeepFM \cite{DeepFM}, which combines FM (factorization machines) and deep learning architecture, play not well, since they solely extracts implicit interactions of short-term behaviors of users without aware of the purchase sequences and the relations of items.
On the other hand, unlike shopping fruits or snacks, users are acting much more reasonably and making decision much more cautiously when selecting a financial commodity.
Therefore, the traditional session-based recommendation methods like newest one SR-GNN\cite{DBLP:SR-GNNShuWu}, have limitations as only using item-purchased sessions but
ignoring the characteristics of users.

As a consequence, we propose a novel recommendation architecture on session-based Graph Neural Network, with the enhancement of learning sophisticated features embeddings of users through DeepFM method as shown in Figure 1.

The main contributions of our works include:
\begin{itemize}
\item We propose a novel session-based recommendation architecture to solve the massive-scale and complicated recommendation scenario with both virtual and real commodities.

\item Our method takes advantages of latent user embeddings and graph-based latent item embeddings, to make an impressive progress on prediction of next-purchase commodity.

\item Through the off-line experiments on the real transactions dataset of \textbf{Jinguanjia}, our model has shown improvement on precision and MRR than other state-of-the-art models.
Additionally, we have completed the low stream test in the live environment and achieved advance than previous methods.
\end{itemize}
 %tailored feed list of recommended items for each user is needed, through analyzing the features of items, users and users interaction with items. This function is named \textbf{Weinituijian} driven by our well-designed recommendation system.

%Many previous works have addressed recommendation algorithms with deep learning methods.
%In\cite{DBLP:journals/corr/abs-1901-08907} , Meituan-Dianping Group advocated a Multi-task feature learning approach
%for knowledge graph enhanced recommendation, which presented a end-to-end deep KG network;
%In \cite{DBLP:journals/corr/abs-1801-01725}, members from Alibaba Group presented achieved recommendation product title compression via a network-based seq2seq approach. In \cite%{DBLP:journals/corr/abs-1711-06632}, another
%group from Alibaba Group proposed a novel recommendation framework through
%attention-based DNN in consist of  the heterogeneous behaviors of users. In \cite{DINalibaba}, a deep learning based model is raised for solving the task of click-through rate prediction,%with a Deep
%Interest Network (DIN) which could adaptively learn the representation of user interests from
%historical behaviors.
%In \cite{Baidu_Huang:2016}, members from Baidu Group

\section{problem formulation}
In session-based recommendation scenario, the main target is to predict which item will purchase next. Here is the formulation of this problem as below:

$s=[v_{s,1},v_{s,2},...,v_{s,n}] $ stands for a purchase session sequence, ordered by timestamps.
$V=[v_1,v_2,...,v_m] $ stands for all unique items involved in all the purchase sessions.
The target of recommendation is to predict the next purchase item $v_{s,n+1}$ in $s$.
In addition, we collect the portraits dataset of users in session sequences, such as characteristics of population, financial behavior, consumption behavior and so on.
For each user, to judge a session-based recommendation model, for each session $s$, the model generates probability vector $\hat{\textbf{y}}$ for all items in $V$, and each value of $\hat{\textbf{y}}$ is the recommendation score of the corresponding item. Thus, the top-K recommendation will select K items from $\hat{\textbf{y}}$ correspondingly.

\section{proposed method}
In this section, we illustrate our method named UBER-GNN (user-based embeddings recommendation on Graph Neural Networks), as shown in Figure 2. At first, we explain the vital step how user-based embeddings are generated. Second, in SECTION 3.2 and 3.3, we construct a GNN from purchase-sessions. Third, we describe the combinations of user embeddings and session embeddings of items. Finally, we present the details of model training.

\subsection{User context based embeddings}
In this section, we explain the extraction user-context based embeddings from the portraits data of users. Inspired by the end-to-end method DeepFM\cite{DeepFM}, we carry out low-order (order-2) feature interactions from Factorization Machines and high-order (above order-3) from DNN. The portraits data of users includes categorical and continuous features. To construct features embeddings, initially each categorical one is represented as a vector of one-hot encoding, and each continuous one is represented as a vector of one-hot encoding after discretization. we regard $\textbf{X}=[e_1,e_2,...,e_n]$ as the input of DeepFM, where $\textbf{X}$ is a $h$-dimensional vector, with $e_i$ being the vector of i-th field and n is the number of fields.
After fed into DeepFM, we can generate user embeddings like:

\begin{equation}\textbf{U} = \sigma(\Upsilon_{(FM)}+\Upsilon_{(DNN)})\end{equation}
where $\Upsilon _{FM}$ is the output of FM part and  $\Upsilon _{DNN}$ is the output of DNN part. $\sigma()$ is a sigmoid function.
\subsubsection{FM part}

The FM part is a factorization machine, which can learn order-1 by addition way and order-2 by inner product way of feature interactions, shown as:
\begin{equation}\Upsilon_{(FM)}= <W,X> + \sum_{j_1}^n\sum_{j_2=j_1+1}^n<k_{i},k_{j}>e_{j_1}\cdot e_{j_2}  \end{equation}
where $<W,X>$ is the linear transformation of $X$ and the other part reflects the order-2 transformation. $k_{i}$,$k_j$ are latent vectors.

\subsubsection{DNN part}
The DNN part is a feed-forward neural network, which can learn high-order feature interactions.
$\alpha^{(0)}$=$\textbf{X}$ is input into DNN and for each layer:
\begin{equation}\alpha^{(i+1)}=\sigma(W^{(i)}\alpha^{(i)}+b^{(i)})\end{equation}
where i is the layer depth and $\sigma$ is an activation function. $\alpha^{(i)}$,$W^{(i)}$,$b^{(i)}$ are the output of previous layer, weight and bias of the layer $i$.
Thus $\Upsilon_{(DNN)}=\sigma(W^{(L)}\alpha^{(L)}+b^{(L)})$, where $L$ is the number of hidden layers of DNN.

\subsection{Session graph representation}
In this section, we introduce a method to represent graph. For each session sequence $s$, there is a directed graph structure $G_s=(V_s,E_s)$, where nodes collection $V_s$ takes each node $v_{s,i} \in V$, edges collection $E_s$ takes each $(v_{s,i-1},v_{s,i})$ in session sequence $s$. $(v_{s,i-1},v_{s,i})$ stands for a user purchase $v_{s,i}$ after $v_{s,i-1}$ in the session $s$. Because some items may be purchased more than once in
a session $s$, we assign each edge with a normalized weight, which is calculated as the occurrence of the edge divided by the outdegree of the start node of the edge.

Besides, the connection matrix $\textbf{A}_s \in \mathbb{R}^{d|V_s|\times2d|V_s|}$ is introduced to represents a session graph $s$ with an unique structure.
$\textbf{A}_s = [\textbf{A}_s^{(out)},\textbf{A}_s^{(in)}]$, which represents weighted connections of outgoing and incoming edges in the session graph respectively. For example, a session graph $G_s$ and the connection matrix $\textbf{A}_s$ is shown in Figure 3, and the corresponding session is $s=[v1,v2,v3,v2,v3,v2,v4]$ having nodes $v_2$ and $v_3$ purchased over once.

 Meanwhile, our model will transform every item $v \in V$ into a same embedding space with a node vector $\textbf{v} \in \mathbb{R}^d$ which is a latent vector learned via graph neural network(\textbf{GNN}). Additionally, each session can be represented by an embedding vector $\textbf{s}$ at SECTION 3.4, which is combined through the node vectors $\textbf{v}$ generated from GNN.

\subsection{Item embeddings via GNN}

\begin{figure}
  \centering
  \includegraphics[scale=0.3]{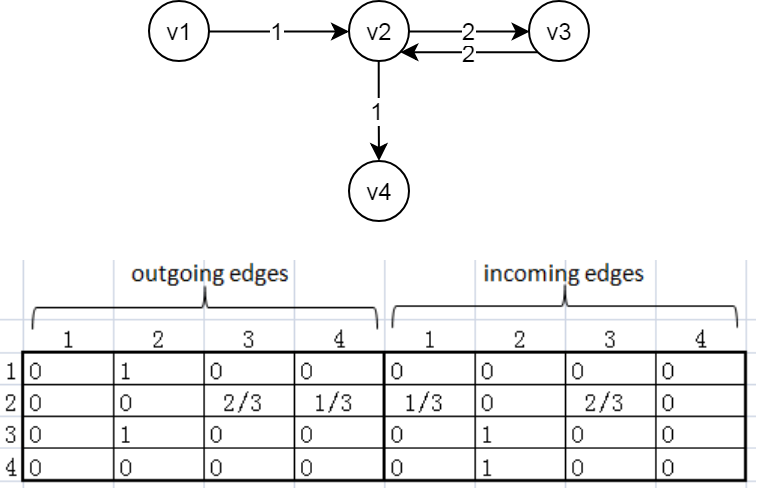}
  \caption{An example of session graph representation with the connection matrix $\textbf{A}_s$}
  \Description{The 1907 Franklin Model D roadster.}
\end{figure}
Thirdly,
The plain vanilla GNN is proposed by \cite{article:Scarselli}, then \cite{DBLP:journals/corr/abs-1511-05493} improved GNN by replacing  propagation model with gated recurrent units(\textbf{GRUs}) and proposed Gated graphed Sequence Neural Networks(GGS-NNs) with  Back-Propagation Through Time(BPTT) to compute gradients. Furthermore, GGS-NNs works are very suitable for session-based recommendation problems, as it can automatically extract features of session graphs with consideration of nodes having many connections.

In addition, there is an analogy can be drawn between the adaptation from GNNs to GGS-NNs, and the adaptation from LSTMs\cite{Hochreiter:1997:LSTM} to GRUs\cite{DBLP:GRU_Cho} in Recursive Neural Networks \cite{SocherEtAl2011:RNN}.
In GGS-NNs, instead of the standard GNN recurrence, new adaptation can improve the long-term
propagation of information across a graph structure.

In the GRU of GGS-NNs, the gated structure is shown like below:

\begin{equation}\textbf{a}_{s,i}^t = \textbf{A}_{s,i:}[\textbf{v}_1^{t-1},...,\textbf{v}_n^{t-1}]^T\textbf{H}+\textbf{b}  \end{equation}
\begin{equation}\textbf{z}_{s,i}^t = \sigma(\textbf{W}_z\textbf{a}_{s,i}^t+\textbf{U}_{z}\textbf{v}_i^{t-1}) \end{equation}
\begin{equation}\textbf{r}_{s,i}^t = \sigma(\textbf{W}_z\textbf{a}_{s,i}^t+\textbf{U}_{r}\textbf{v}_i^{t-1}) \end{equation}
\begin{equation}\tilde{\textbf{v}}_{s,i}^t = tanh(\textbf{W}_o\textbf{a}_{s,i}^t+\textbf{U}_{o}(\textbf{r}_{s,i}^t\odot\textbf{v}_i^{t-1}) \end{equation}
\begin{equation}\textbf{v}_{s,i}^t = (1-\textbf{z}_{s,i}^t )\odot\textbf{v}_i^{t-1}+\textbf{z}_{s,i}^t\odot\tilde{\textbf{v}}_{s,i}^t  \end{equation}
Where $\textbf{z}_{s,i}^t$ and $\textbf{r}_{s,i}^t$ are the update gate and reset gate, $\textbf{v}_{s,i}^t$  and $\tilde{\textbf{v}}_{s,i}^t$ are the activation and the candidate activation.
In equation(4), $\textbf{A}_{s,i:} \in \mathbb{R}^{(1\times2n)}$ is the concatenation of two columns of $\textbf{v}_{s,i}$ in $A_s$; $[\textbf{v}_1^{t-1},...,\textbf{v}_n^{t-1}]^T$ is the list of node vectors in session $s$ and $n$ represents the total number of total nodes in $V_s$;$\textbf{H}\in \mathbb{R}^{(d\times2d)}$ and $\textbf{b}\in \mathbb{R}^{(1\times2d)}$ are  hyper-parameters matrices where d is pre-set as the dimension of item embeddings $v_i$.
In equation (5) and (6), operator $\sigma()$ is the sigmoid function, and $\textbf{W}_z$ $\textbf{W}_z$ $\textbf{U}_{z}$  $\textbf{U}_{r}$ are hyper-parameters matrices.
In equation (7) and (8), operator $\odot$ is the element-wise multiplication operator and $\textbf{W}_o$ $\textbf{U}_{o}$ are hyper-parameters matrices.

To better illustrate, here we explain how GRU generate vector of item embeddings.
GGS-NN can proceed nodes of session graph $G_s$ at the same time. In each GRU, equation (4) is used for information
propagation between different nodes of the session graph construction $A_s$ of $G_s$. Specifically, at first,the GRU extracts the latent
vectors of neighborhoods and feeds them as input into the neighbor GRUs. Second, the update gate of equation (5) and reset
gate of equation (6), utilize the sigmoid function to decide what information to be preserved or discarded
respectively. Third, equation (7) constructs the candidate state by
the current state, the reset gate and the previous gate. Finally, the final state of equation (8) is the combination of
the previous state and the candidate state, with the consideration of the update gate. As a consequence, after all nodes in session
graphs is processed by GGS-NN until convergence, we can obtain the final node vectors of all item embeddings.

\subsection{Session embeddings with attention to user context}
Regardless of user static features, previous session-based recommendation methods only focus on session sequences.
On the contrary, to better predict the user's next purchase item, we address an attention mechanism to
combine users context as user embedding vectors and purchase sessions as item embedding vectors.
In this section, we illustrate how to enhance session-based method with attention to user context.
\begin{equation}\lambda_i=\textbf{a}^T\sigma(\textbf{W}_1 [\textbf{U};\textbf{S}_c]+\textbf{W}_2\textbf{v}_i +\textbf{C}) \end{equation}
\begin{equation}\textbf{S}_g=\sum_{i=1}^N\lambda_i \textbf{v}_i\end{equation}
where $\textbf{U} \in  \mathbb{R}^M$  is the output from SECTION 3.1 which is unique for each user.
$\textbf{a}\in\mathbb{R}^d$, $\textbf{W}_1 \in\mathbb{R}^{(d\times (M+d))}$,$\textbf{W}_2\in \mathbb{R}^{(d\times d)}$,$\textbf{C}\in \mathbb{R}^{d}$
as hyper-parameters, control the weights of embeddings. In SECTION 3.3, we generate the vectors of all nodes after feed all session graphs into
GGS-NNs. Then in equation (10), instead of using $s=[v_{s,1},v_{s,2},...,v_{s,n}]$ to represent user purchase session $s$, we extract essential
latent information as $s_g$ by aggregating all nodes vectors. Moreover, to consider the influence of last-time purchase, we assign $s_c=v_n$.
therefore, we apply a hybrid combination $s_h$ through the concatenation of current purchase interest, global purchase preference and user's latent
context.

\begin{equation}\textbf{S}_h=\textbf{W}_3[\textbf{S}_c;\textbf{S}_g;\textbf{U}]\end{equation}
where matrix $\textbf{W}_3 \in \mathbb{R}^{d \times (2d+M)}$ compresses three embedding vectors into latent space $\mathbb{R}^d$.

Furthermore, to emphasize the effect of different attention approaches, we assign adaptation to $\textbf{S}_g$.

\noindent{\verb|(1)|}  global embedding with average pooling, equation (10) is changed as:
\begin{equation}\textbf{S}_g=\sum_{i=1}^N\frac{1}{N}\textbf{v}_i\end{equation}
\noindent{\verb|(2)|}  global embedding with attention mechanism that only considers local embedding, equation (9) is changed as:
\begin{equation}\lambda_i=\textbf{a}^T\sigma(\textbf{W}_1 \textbf{S}_c+\textbf{W}_2\textbf{v}_i +\textbf{C}) \end{equation}
\noindent{\verb|(3)|}  global embedding with attention mechanism that only considers user embedding, equation (9) is changed as:
\begin{equation}\lambda_i=\textbf{a}^T\sigma(\textbf{W}_1 \textbf{U}+\textbf{W}_2\textbf{v}_i +\textbf{C}) \end{equation}

All the details of comparison is evaluated in SECTION 4.5.
\subsection{Model training}
With the compression process, we obtain each session latent vector, then we can compute the recommended score-value $\hat{\omega_i}$ for each purchase
item $v_i \in \textbf{V}$ by times item vector $v_i$ and session vector $s_h$, which is defined as:
\begin{equation}\hat{\omega_i}=\textbf{S}_h^T \textbf{v}_i\end{equation}

Then the score vector ${\boldsymbol{\hat\omega} \in \mathbb{R}^{N}}$ where N stands for the total number of items is fed into softmax function to
get the output probabilities:
\begin{equation}\hat{\textbf{y}}=softmax(\hat{\boldsymbol{\omega}})\end{equation}
where  $\hat{\textbf{y}} \in \mathbb{R}^{N}$ denotes the probability of items which is the next purchase one in session $s$.

Finally, we take cross-entropy as the loss function of the prediction and the ground truth for each session $s$. It is shown as:
\begin{equation} \boldsymbol{L(\textbf{y})}=-\sum_{i=1}^N\textbf{y}_i log(\hat{\textbf{y}}_i)+(1-\textbf{y}_i)log(1-\hat{\textbf{y}}_i) \end{equation}

Accordingly, to train the whole model, we use the Back-Propagation Through Time(BPTT) algorithm.
\section{Experimental result}

\subsection{Datasets description}
\begin{table}
\centering
\addtolength{\tabcolsep}{-4pt}
\caption{Details of datasets in our experiments}
\begin{tabular}{|c|c|c|} \hline
Dataset & financial-include & financial-exclude \\
\hline
\hline
\# of commodities &82,126&65,623\\ \hline
\# of users &24,022&21,921\\ \hline
\# of transactions &753,960&473,495\\ \hline
\# of sessions &91,433&76,816\\ \hline
Avg length of session&8.246&6.164\\ \hline
\end{tabular}
\end{table}
We implement and evaluate our method on real-world transaction data of Ping An \textbf{Jinguanjia}, which owns sufficient, multidimensional portraits of users to
generate precise and comprehensive embeddings of user features.

Datasets have two major parts; First is the portraits data of users, including 306 features (212 categorical ones and 94 continuous ones);
Second is the purchase session data of users, within 12 months selected from March 2018 to March 2019.
Besides, to fair compare, we extract sessions data of length $>$ 1 and filter out items appearing $<$ 5 times from the initial datasets. And to generate the labels, we
split the purchase sessions. For example, a purchase session $s=[v_{s,1},v_{s,2},...,v_{s,n}]$ can be extracted sequences and labels like, $([v_{s,1}],v_{s,2})$,$([v_{s,1},v_{s,2}],v_{s,3})$,
..., $([v_{s,1},...,v_{s,n-1}],v_{s,n})$, where $v_{s,n}$ is next-purchase item. Furthermore, considering \textbf{Jinguanjia} as a complex E-Commerce platform supplies not only real commodities
but also virtual financial commodities, we select sessions without financial commodities as financial-exclude dataset.

The statistics of datasets are summarized in Table 1.
\subsection{Baseline methods}
For better comparison, we choose baseline methods as follows:
\begin{itemize}
\item {\verb|BPR|}: \cite{Rendle:2010}optimized a pairwise ranking objective function via stochastic gradient descent.
\item {\verb|DeepFM|}: \cite{DeepFM}  combined FM and deep learning architecture to learn implicit user behaviors for CTR prediction.
\item {\verb|GRU4REC|}:\cite{DBLP:journals/corr/HidasiKBT15} used RNNs to model user sequences for the session-based recommendation.
\item {\verb|SRGNN|}:\cite{DBLP:SR-GNNShuWu} used GNNs to model session sequences as graph structured data.
\end{itemize}
\subsection{Evaluation Metrics}

\noindent\textbf{P@20}(Precision) stands for the proportion of correctly recommended items among the top-20 items.

\noindent\textbf{MRR@20}(Mean Reciprocal Rank) stands for a statistic measure for evaluating average accuracy of ranking in top-20. The query of a reciprocal rank of top-20 is the multiplicative inverse of the rank of the first correct answer: 1 for first place, 1${/}$2 for second place, 1${/}$3 for third place and 0 for exceeding top-20.
\subsection{Parameters settings}
In our experiments, validation set is a random 20 percent subset of the training set. All parameters are initialized using a Gaussian distribution with a mean of zero and standard deviation of 0.1.
In our methods, we set the hyper-parameters as following: the hidden size and the batch size is set to 200 and 32 respectively.
The mini-batch Adam optimizer is utilized to optimize parameters, where the initial learning rate is set to 0.1 and will exponential decay to 0.01 after every 10 epoches. Moreover, the L2 penalty is set to 1e-5 to get better performance.

\subsection{Results analysis}

\noindent{\bf Comparison with baseline methods}:

In order to evaluate the overall performance of our proposed model, we compare it with other state-of-art session-based recommendation methods and classical CTR prediction method DeepFM. The overall performance in terms of P@20 and MRR@20 is shown in Table 2.

In our proposed UBER-GNN model, it jointly utilizes both graph-structure data aggregated by session sequences and user-based classical-structure data. Thus our model consider user long-term static latent characteristics and preferences as well as their dynamic latent interests. According to the experiments, our proposed UBER-GNN model achieves the best performance on both two datasets in terms of Precision@20 and MRR@20.

Regarding traditional algorithms like BPR,the performance is relatively poor. Such simple methods make recommendations solely based on history items, which is problematic in session-based recommendation scenarios. As well as DeepFM, it's not suitable when predicting next-purchase. Likewise, Short/Long-term memory models, like GRU4REC, use recurrent units to capture a user's global interest. Such method explicitly model the user's global behavior preferences. While graph neural network based models, like SR-GNN, transfer session history sequences into graph-structured data and utilize gated graph neural network to update item embeddings. However, their performances are still inferior to that of our proposed UBER-GNN Model.

Compared with the state-of-art models like SR-GNN and GRU4REC, our model firstly considers user-based classical-structure data to better represent user long-term latent characteristics and preferences, and further models transitions between items into graph that can capture more complex and implicit connections between recent behaviors. Whereas in GRU4REC and SR-GNN, they explicitly model item history and obtain user representations through sequences, losing sight of the characteristics information of users.
\begin{table}
\vspace{0.25em}
\centering
\small
\addtolength{\tabcolsep}{-1pt}
\caption{The performance of UBER-GNN compared with other baseline methods}
\begin{tabular}{|c|c c|c c|}
\hline
\hline
\raisebox{-2ex}[0cm][0cm]{Method} & \multicolumn{2}{p{90pt}|}{\centering{financial-include}} &  \multicolumn{2}{p{90pt}|}{\centering{financial-exclude}}\\
\cline{2-5}{}& P@20(\%) & MRR@20(\%) & P@20(\%) & MRR@20(\%) \\
\hline
\hline
\textbf{BPR} &51.43&12.48&53.77&13.24\\ \hline
\textbf{DeepFM} &65.03&14.41&63.89&13.75\\ \hline
\textbf{GRU4REC} &68.48&20.09&65.59&18.84\\ \hline
\textbf{SRGNN} &74.64&24.47&75.62&25.01\\ \hline
\textbf{Ours}&77.91&26.66&78.04&25.75\\ \hline \hline
\end{tabular}
\end{table}

\noindent{\bf Comparison with variants on session embedding strategy}:

We compare the session embedding strategy with following different attention approaches:

{\verb|(1)|}  global embedding with average pooling.

{\verb|(2)|}  global embedding with attention mechanism that only considers local embedding.

{\verb|(3)|}  global embedding with attention mechanism that only considers user embedding.

{\verb|(4)|}  global embedding with attention mechanism that both considers user embedding and local embedding.

The result of methods with four different strategies on both two datasets are given in Table 3.

Firstly, the result is shown that attention mechanisms are useful in extracting significant behaviors from history sequences. Average pooling strategy may not be adaptive for the sequence scenario due to uncertain noisy behaviors.

Furthermore, hybrid attention embedding strategy considering both static user embedding and dynamic local embedding achieves best results on both two datasets. It validates the importance of explicitly incorporating dynamic latent features and static latent features. In additional, it supports that both static user features and dynamic local features are crucial for session-based recommendation.

\begin{table}
\vspace{0.25em}
\centering
\small
\addtolength{\tabcolsep}{-1pt}
\caption{The performance of UBER-GNN compared with its four variants }
\begin{tabular}{|c|c c|c c|}
\hline
\hline
\raisebox{-2ex}[0cm][0cm]{Method} & \multicolumn{2}{p{100pt}|}{\centering{financial-include}} &  \multicolumn{2}{p{100pt}|}{\centering{financial-exclude}}\\
\cline{2-5}{}& P@20(\%) & MRR@20(\%) & P@20(\%) & MRR@20(\%) \\
\hline
\hline
\textbf{V1} &75.21&24.92&75.42&24.98\\ \hline
\textbf{V2} &77.01&26.22&75.83&25.34\\ \hline
\textbf{V3} &76.59&25.87&77.05&25.56\\ \hline
\textbf{V4} &77.91&26.66&78.04&25.75\\ \hline \hline
\end{tabular}
\end{table}

\section{Conclusions}
In this paper, we propose a novel architecture for session-based recommendation that takes long-term preference and dynamic interests into consideration.
The proposed method UBER-GNN not only takes advantage of structured data to generate long-term user preferences, but also transfers session sequences into graphs to generate graph-based dynamic interests.
In addition, it develops an attention strategy to ensemble long-term preferences and dynamic interests to better predict users' next actions.
Extensive experiments conducted on real Ping An scenario show that UBER-GNN outperforms the state-of-the-art session-based recommendation methods.
\begin{verbatim}

\end{verbatim}

%
% The next two lines define the bibliography style to be used, and the bibliography file.
\bibliographystyle{ACM-Reference-Format}
\bibstyle{acmauthoryear}
\scriptsize
\bibliography{UBER-GNN}

%
% If your work has an appendix, this is the place to put it.

\end{document}